\documentclass{PoS}

\title{Quark masses in two-flavor QCD }

\ShortTitle{Quark masses in two-flavor QCD }

\author{\speaker{Michael Creutz} \thanks{ I am grateful to the award I
    received from Alexander von Humboldt Foundation to visit the
    University of Mainz and which partially supported my attendence to
    this meeting.  This manuscript has been authored under contract
    number DE-AC02-98CH10886 with the U.S.~Department of Energy.
    Accordingly, the U.S. Government retains a non-exclusive,
    royalty-free license to publish or reproduce the published form of
    this contribution, or allow others to do so, for U.S.~Government
    purposes.  }\\ Brookhaven National Laboratory\\ E-mail:
  \email{creutz@bnl.gov}}

\abstract{Considered as a function of the quark mases, two-flavor QCD
  depends on three parameters, including one that is CP violating. As
  the masses vary to unphysical values, regions of both first- and
  second-order phase transitions are expected.  For non-degenerate
  quarks, non-perturbative effects leave individual quark mass ratios
  with a renormalization scheme dependence.  This complicates matching
  lattice results with perturbative schemes and clarifies the
  tautology with attacking the strong CP problem via a vanishing up
  quark mass.}

\FullConference{International Workshop on QCD Green's Functions,
Confinement and Phenomenology\\
         5-9 September 2011\\
         Trento, Italy}

\input epsf

\begin{document}

\section{Introduction}

At the previous meeting in this series \cite{Creutz:2009zq} I
discussed the fascinating physics arising from the interplay of the
three ways chiral symmetry is broken in QCD.  These are (1) the
spontaneous breaking responsible for the lightness of pions, (2) the
breaking of the singlet axial $U(1)$ symmetry by the anomaly, and (3)
the explicit breaking of chiral symmetry by the quark masses.  For
simplicity, that discussion was restricted to degenerate quarks.  Here
I move on to some interesting generalizations that occur when the
quarks are no longer degenerate.  Since the number of parameters grows
with the number of flavors, I concentrate here on the two flavor
theory and consider what happens when the quark masses are varied from
their physical values.  As a function of the these parameters a rather
intricate phase diagram emerges, displaying both first and second
order phase transitions.  Much of this talk is adapted from the more
detailed treatments in Refs.~\cite{Creutz:2010ts} and
\cite{Creutz:2011hy}.

To begin, let me remind you of the expected behavior of two flavor QCD
in the limit of massless quarks.  Because of confinement and
dimensional transmutation, this theory should possess several massive
states, including the proton, neutron, eta prime, and glueballs.  In
addition, spontaneous chiral symmetry breaking should give rise to
three massless pions as Goldstone bosons.  In this picture both the
eta prime and the neutral pion are composites of distinct mixtures of
$\overline u u$ and $\overline d d$ quarks.  The eta prime, defined as
the lightest isosinglet pseudoscalar, also has a contribution from
purely gluonic constituents.  The latter are related to the anomaly
and the fact that the $\pi_0$ and the $\eta^\prime$ are not
degenerate.

\begin{figure}
\begin{centering}
\includegraphics[width=.45\textwidth]{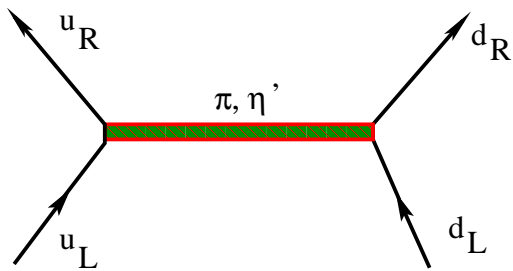}
\caption{Because of the anomaly, spin-flip scattering of massless up
 and down quarks does not vanish.}
\label{fig1}
\end{centering}
\end{figure}

In this theory, consider a hypothetical quark-quark scattering
experiment, as sketched in Fig.~\ref{fig1}.  This represents spin flip
scattering of an up quark against a down quark.  Exchanges of both the
neutral pion and the eta prime can contribute to this process.
Because these particles are non-degenerate, their contributions cannot
cancel.  Therefore, the spin-flip four point function does not vanish.
Were it not for the anomaly, the two exchanges could cancel.

\begin{figure}
\begin{centering}
\includegraphics[width=.45\textwidth]{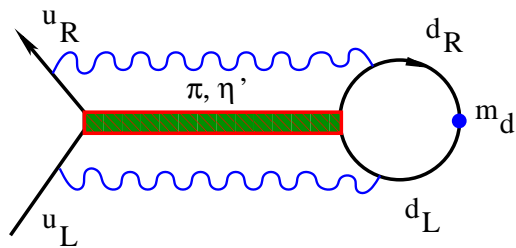}
\caption{A small down quark mass induces an additive shift in the up
  quark mass through pseudoscalar meson exchange.
}
\label{fig2}
\end{centering}
\end{figure}

Now turn on a small down quark mass.  Take the diagram in
Fig.~\ref{fig1} and close the down quark lines into a loop with a mass
insertion as shown in Fig.~\ref{fig2}.  This provides a mechanism for
mixing the left and right handed up quark, {\it i.e.} the up quark
develops an effective mass.  Starting with a vanishing up quark mass,
the mass ration $m_u\over m_d$ becomes renormalized by
non-perturbative effects.  Except in the isospin limit, quark mass
ratios will not be renormalization group invariant.  Since lattice
gauge simulations include all non-perturbative physics, this effect is
automatically present in such calculations.

This trivial observation is rather old and is often discussed in terms
of instanton physics
\cite{Georgi:1981be,Banks:1994yg,Creutz:2003cj,Creutz:2003xc}.  Note
that only the last of these references appears in a refereed journal,
more than twenty years after the first.  This is a consequence of an
intense consternation from the perturbative community based on the
lore that spin flip processes are suppressed in the massless limit.

This renormalization of quark mass ratios is an effect not seen in
conventional perturbative schemes, such as $\overline{MS}$.  The
consequences have therefore been quite controversial.  Mass
renormalization is not flavor blind, and a mass independent
renormalization scheme is problematic.  The $\overline{MS}$ scheme is
not a complete regulator since it ignores such non-perturbative
effects.  The crucial conclusion here is that when the up and down
quarks are not degenerate, then attempts to match masses obtained from
lattice calculations with perturbative results are inherently
meaningless.  I am not criticizing $\overline{MS}$ as a perturbative
regulator; rather, the lattice and perturbative calculations involve
different physics and should not be compared.

\section{Specific critiques}

The above observations raise frequent objections.  At the simplest
level, one might try to claim that the concept of $m=0$ corresponds to
the bare mass rather than some running quantity.  The problem with
this is that the bare quark masses always vanish.  The renormalization
group tells us that as one approaches the continuum limit
\begin{equation}
m_0 \propto g_0^{\gamma_0/\beta_0} (1+O(g_o^2))
\end{equation} 
with the known coefficients
\begin{equation}
\matrix{
&\beta_0=&{11-2n_f/3 \over (4\pi)^2}\cr
&\gamma_0=&{8\over (4 \pi)^2}.\cr
}
\end{equation}
The asymptotic freedom result that the bare coupling $g_0$ goes to
zero in the continuum limit then immediately implies $m_0\rightarrow
0$.  To talk about quark masses as non-vanishing quantities, it is
necessary to define them using some finite scale.

A more sophisticated complaint is that one has the option to use a
mass independent regularization scheme.  In the renormalization group
equation for the mass
\begin{equation}
a {dm_i\over da}=\gamma(g) m_i\ 
\end{equation}
only the leading perturbative term in $\gamma(g)$ is scheme
independent.  If one requires that $\gamma(g)$ is independent of any
of the quark masses, then one automatically obtains
\begin{equation}
{m_i\over m_j}=
\hbox{constant}.
\end{equation}
Indeed, such a regularization is technically allowed, but it hides
the above off-diagonal $m_d$ effect on $m_u$.  There is no guarantee
that quark mass ratios are independent of scheme, and the lattice, as
usually implemented, is itself not a mass independent scheme.  This
makes it quite obscure how to do a matching with lattice results.  To
be more specific, when $m_u$ is different from $m_d$, isospin is
broken explicitly and the charged pion mass differs from that of its
neutral partner.  A straightforward effective Lagrangian analysis
relates the ratio of pion masses to the quark masses with the result
\begin{equation}
{m_{\pi^0}^2\over m_{\pi^\pm}^2} =1-O\left({(m_u-m_d)^2\over
(m_u+m_d) \Lambda_{qcd}}\right)
\end{equation}
As an immediate consequence, if one holds the quark masses fixed, then
the physical hadronic mass ratios will be scale dependent.
Conversely, if one holds the hadron mass ratios fixed, as usually done
in lattice simulations, then the quark mass ratios must be scale
dependent.

At this point advocates of the matching process frequently suggest
doing the comparison at some high energy, say 100 GeV, where
instantons are exponentially suppressed and irrelevant.  This does not
resolve the issue for several reasons.  First the lattice simulations
are not done at such small scales and the instanton effects must be
included.  Furthermore, the asymptotic freedom result
\begin{equation}
1/g^2\sim \log(\mu) \sim \log(1/a)
\end{equation}
shows that the exponential suppression in $1/g^2$ is actually only a
power law suppression in the scale.  One can easily estimate the size
of these effects from the renormalization group, which tells us that
\begin{equation}
m_{\eta^\prime} \propto {1\over a}{ e^{-1/(2\beta_0 g^2)}
g^{-\beta_1/\beta_0^2}} \not \rightarrow 0.
\label{etaprime}
\end{equation}
The uncertainty in the up quark mass is proportional to this mass as
well as being proportional to $m_d-m_u$.  Thus the expected order of
the up quark mass shift at a scale of a few GeV is
\begin{equation}
\Delta m_u(\mu) \sim {\scriptstyle (
m_{\eta^\prime}-m_{\pi_0})\ (m_d-m_u) \over \Lambda_{qcd}}
= O(1\ \hbox{MeV}).
\end{equation}
This is a number comparable in size to the quoted lattice masses
\cite{Bazavov:2009bb,Blum:2010ym,Durr:2010vn}.

In this context it is important to note that the exponent in
Eq.~(\ref{etaprime}), ${\scriptstyle 8\pi^2\over (11-2n_f/3) g^2}$, is
considerably smaller than the classical instanton action $
{\scriptstyle 8\pi^2\over g^2}.$ This emphasizes that the relevant
topological excitations need to be considered above the quantum, not
the classical vacuum.  Calculations based on the classical instanton
solution strongly underestimate these effects.  The renormalization
group gives the correct suppression.

\section{General masses in two flavor QCD}

I now arrive at the main topic of this talk, the most general mass
parameters for two flavor QCD.  A mass term should be a
dimension-three Hermitean quadratic form in the quark fields.  As well
it should be Lorentz invariant and electrically neutral.  Based on
these criteria, the most general expression is
\begin{equation}
m_1\ \overline\psi\psi+
m_2\ \overline\psi\tau_3\psi+
im_3\ \overline\psi\gamma_5\psi+
im_4\ \overline\psi\tau_3\gamma_5\psi.
\end{equation}
Conventionally one might refer to these four terms with $m_1$
representing the average quark mass, $m_2$ the up-down mass
difference, and $m_3$ a possible CP violating term related to the
Theta parameter.  Finally $m_4$ represents what is sometimes called a
``twisted mass.''

These four mass parameters are not independent.  Consider a flavored
chiral rotation of form
$\psi\rightarrow e^{i\theta\tau_3\gamma_5}\psi$.  Under this the
various quadratic forms transform as
\begin{equation}
\matrix{
&\overline\psi\psi\hfill&\rightarrow &
\cos(\theta)&\overline\psi\psi\hfill
&+&\sin(\theta)&i\overline\psi\gamma_5\tau_3\psi\hfill\cr
&\overline\psi\tau_3\psi\hfill&\rightarrow &
\cos(\theta)&\overline\psi\tau_3\psi\hfill
&+&\sin(\theta)&i\overline\psi\gamma_5\psi\hfill\cr
&i\overline\psi\gamma_5\psi\hfill&\rightarrow &
\cos(\theta)&i\overline\psi\gamma_5\psi\hfill
&-&\sin(\theta)&\overline\psi\tau_3\psi\hfill\cr
&i\overline\psi\tau_3\gamma_5\psi\hfill&\rightarrow& 
\cos(\theta)&i\overline\psi\tau_3\gamma_5\psi\hfill
&-&\sin(\theta)&\overline\psi\psi\hfill\cr
}
\end{equation}
This rotation mixes $m_1\leftrightarrow m_4$ and $m_2\leftrightarrow
m_3$.  What is essentially a change of variables allows one to select
any one of the $m_i$ to vanish and a second to be positive.

The conventional choice is to take $m_4=0$ and then use $m_1>0$
for the average quark mass and $m_2$ for the quark mass difference.
The CP odd term proportional to $m_3$ is related to the Theta
parameter and will be discussed further momentarily.

An alternative choice is to select $m_1=0$ and use $m_4>0$ as the
average quark mass.  Then the quark mass difference moves to the $m_3$
term and $m_2$ encodes the CP violation.  This is the choice used for
``twisted mass'' lattice simulations.  The primary motivation lies
with certain lattice artifacts which depend on the twist.  These are
minimized with this choice \cite{Frezzotti:2003ni,Munster:2004wt}.

It is important to recognize that the choice between these options is
purely a convention and the continuum physics is equivalent between
them.  For the following discussion I adopt the first and more
familiar approach with $m_4=0$.

A crucial aspect of this theory is how the anomaly prevents rotations
between $m_1\overline \psi \psi$ and $im_3\overline\psi\gamma_5\psi$.
Such would follow from a hypothetical variable change
\begin{equation}
\psi\rightarrow e^{i\theta\gamma_5}\psi.
\end{equation}
This however is not a valid symmetry
\cite{Adler:1969gk,Adler:1969er,Bell:1969ts,Jackiw:1999bc} because it
changes the fermion measure
\begin{equation}
d\psi\rightarrow e^{i\theta{\rm Tr}\gamma_5}d\psi.
\end{equation}
The issue, as nicely elucidated by Fujikawa \cite{Fujikawa:1979ay}, is
that in any regulated theory $\gamma_5$ cannot remain traceless.  For
example, consider a cutoff $\Lambda$ and regulate the theory
suppressing large eigenvalues of the
Dirac operator $D$.  The index theorem gives the result
\begin{equation}
{\rm Tr}\gamma_5e^{D^2/\Lambda^2}=\nu
\end{equation}
where $\nu$ is the winding number of the gauge field configuration
under consideration.  Thus the above rotation will introduce a factor
of $\exp(i\theta \nu)$ into the path integral and thereby change the
value of the QCD Theta parameter.  Actually, the above rotation allows
one to move any Theta parameter from the gauge action into the mass
terms.  For the following, assume that this has been done.  After
this, all three mass parameters are both relevant and independent.

\section{The strong CP problem} 

Experimentally the strong interactions preserve CP symmetry to high
accuracy.  This would not be the case if $m_3$ were substantial.
Indeed, only the two parameters $m_1$ and $m_2$ seem to be needed.
The strong CP problem asks why is $m_3$ so small?

This issue arises because of the possible unification of interactions.
The weak interactions are known to violate CP; so, when the
interactions separate as one goes down in energy, why is it that some
residue of the CP violation doesn't remain in a non-vanishing $m_3$.

One trivial ``solution'' is that there is no unification.  One could
consider the strong interactions on their own and impose CP symmetry
from the outset.  In this picture the weak interactions only come in
as a small perturbation and do not directly affect the Theta angle.

Another approach couples a new dynamical field directly to 
$i\overline\psi\gamma_5\psi$.  In this case $m_3$ becomes a dynamical
quantity and can relax naturally to zero.  This requires a new
particle corresponding to this field, although its coupling is not
determined and could be small.  This is the ``axion'' approach.

It is sometimes proposed that the strong CP problem could be solved by
having the up quark mass vanish.  However the above formalism should
clarify why this is not a sensible approach.  In terms of the three
mass variables, one could define the up quark mass as
\begin{equation}
m_u\equiv m_1+m_2+im_3.
\label{mup}
\end{equation}
The problem is that $m_1$, $m_2$, and $m_3$ are independent parameters
with different symmetry properties.  The parameter $m_1$ represents an
isosinglet mass contribution while $m_2$ multiplies an isovector
quantity.  It is only the parameter $m_3$ which is CP violating.  And
the discussion in the introduction showed that $m_1+m_2=0$ is a scale
and scheme dependent statement.  So while it may be true that setting
$m_u$ from Eq.~(\ref{mup}) to zero would imply $m_3=0$, this could be
regarded as ``not even wrong.''

The basic issue with forcing the up quark mass to zero is that it
involves going to polar coordinates with an unnatural origin.  In a
formal sense one can connect the three mass parameters above with the
more conventional set $\{m_u,m_d,\Theta\}$ via the relations
\begin{equation}\matrix{
m_u=m_1+m_2+im_3,\cr
m_d=m_1-m_2+im_3,\cr
\cr
e^{i\Theta}={m_1^2-m_2^2-m_3^2+2im_1m_3\over
\sqrt{m_1^4+m_2^4+m_3^4+2m_1^2m_3^2+2m_2^2m_3^2-2m_1^2m_2^2
}}.\cr}
\end{equation}
The mixing discussed in the introduction shows that this choice of
parameters, including $\Theta$, is in general scale and scheme
dependent.

\section{The phase diagram}

Taking the mass parameters away from their physical values uncovers a
rather rich phase diagram.  This follows from a simple linear sigma
model analysis.  For this, consider the composite scalar fields
\begin{equation}
\matrix{
&\sigma\propto \overline\psi\psi,
&\vec\pi\propto i\overline\psi\gamma_5\vec\tau\psi,
&\eta\propto i\overline\psi\gamma_5\psi,
&{\vec a_0}\propto \overline\psi\vec\tau\psi. \cr
}
\end{equation}
These provide a model for the two flavor chiral symmetry via an
effective potential
\begin{equation}
\matrix{
V=&\lambda(\sigma^2+\vec\pi^2-v^2)^2-{m_1}\sigma-{m_2}{a_0}_3
-{m_3}\eta\cr
&+{\alpha ({\eta}^2+{\vec a_0}^2)
{-}\beta (\eta\sigma+ {\vec a_0} \cdot \vec \pi)^2}.\cr
}
\end{equation}
The first term, proportional to $\lambda$, is the conventional ``wine
bottle'' or ``Mexican hat'' frequently used to describe spontaneous
symmetry breaking.  The parameters $\alpha$ and $\beta$ can be thought
of as ``low energy constants'' that couple $(\sigma,\vec\pi)$ with
$(\eta,\vec a_0)$.  These combinations rotate similarly under flavored
chiral rotations; so, these constants preserve the chiral symmetry of
the massless theory.  Here the $\alpha$ term serves to give a mass to
the $\eta$ and $\vec a_0$.  The square appearing in the $\beta$ term
is inserted so the basic potential still preserves parity.  The sign of
this term is selected so that $m_{\eta} < m_{\vec a_0}$.

\begin{figure}
\begin{centering}
\includegraphics[width=.5\textwidth]{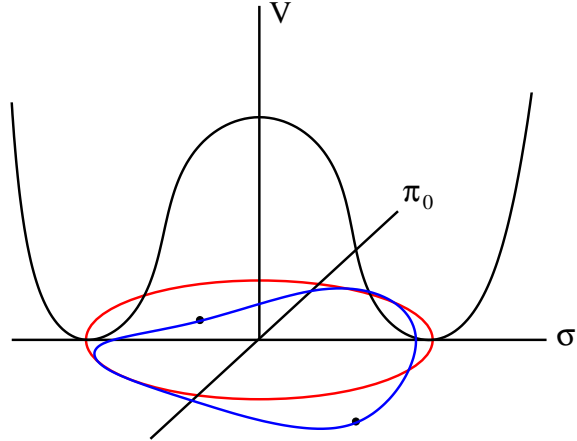}
\caption{The parameters $m_2$ and $m_3$ warp the Mexican hat downward
in a direction determined by their relative size.
}
\label{fig3}
\end{centering}
\end{figure}

The three mass terms break the chiral symmetry in slightly different
ways.  The $m_1$ term serves to tilt this potential and generally
selects a unique minimum.  The effects of the $m_2$ and $m_3$ terms
are more subtle since they do not directly couple to the $\pi$ or
$\sigma$ fields.  With $m_2$ ($m_3$) present the field ${a_0}_3$
($\eta$) will be driven to have an expectation value.  This will feed
back through the $\beta$ term to give a quadratic warping of the
Mexican hat.  This warping will be downward in the $\pi_0$ ($\sigma$)
direction.  With both terms present, this warping will be in some
intermediate direction, as shown in Fig.~\ref{fig3}.  When $m_1$ is
absent, this warping leaves two possible minima into which the vacuum
can settle.  Turning on a small $m_1$, the resulting tilt will select
one or the other as the true vacuum.  This results in a generic first
order transition occuring when $m_1$ changes sign.

\begin{figure}
\begin{centering}
\includegraphics[width=.5\textwidth]{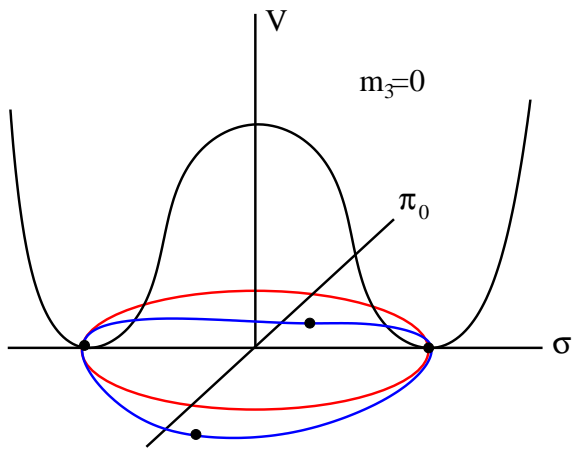}
\caption{At $m_3=0$ the warping is orthogonal to the sigma direction
  and a small $m_1$ term does not select a unique minimum.
}
\label{fig4}
\end{centering}
\end{figure}

\begin{figure}
\begin{centering}
\includegraphics[width=.6\textwidth]{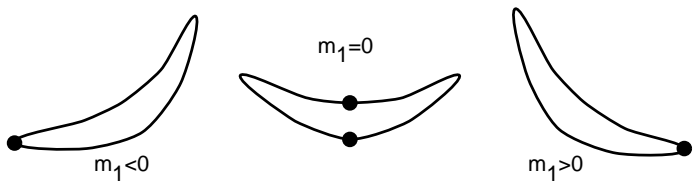}
\caption{As $m_1$ varies at $m_3=0$ there should be a point where a
  single minimum splits into two.
}
\label{fig5}
\end{centering}
\end{figure}

A special case occurs when $m_3=0$ and $m_2\ne 0$.  Then the warping
is downward in the $\pi_0$ direction and $m_1$ does not distinguish
between the two minima, as sketched in Fig.~\ref{fig4}.  In this
situation there will be some intermediate value of $m_1$ where a
single minimum at large tilt splits into two minima with an
expectation value for the neutral pion field.  This is sketched in
Fig.~\ref{fig5}.  At this critical point one expects an Ising-like
behavior.  Here the square of the neutral pion mass passes through
zero and gives rise to a pion condensate.  As the pion is CP odd, this
represents a spontaneous breaking of CP symmetry.

Note that this Ising-like transition at $m_3=0$, $|m_1|<|m_2|$ occurs
with both $m_u$ and $m_d$ non-vanishing, although they are of opposite
sign.  This represents a situation where there is a diverging
correlation length and corresponding long distance physics occuring
without the presence of any small eigenvalues for the Dirac operator.

Conversely, the overall picture indicates no special behavior at
$m_u=0$ when $m_d\ne 0$.  In this case there is no important long
distance physics despite the possibility of small Dirac eigenvalues.
These facts are the seed of many controversies, including the
connection between the strong CP problem and $m_u=0$
\cite{Creutz:2003xc}, the issue of whether topological susceptibility
is a physical observable \cite{Creutz:2010ec}, and the failure of the
rooting process for staggered fermions \cite{Creutz:2007yg}.

\begin{figure}
\begin{centering}
\includegraphics[width=.7\textwidth]{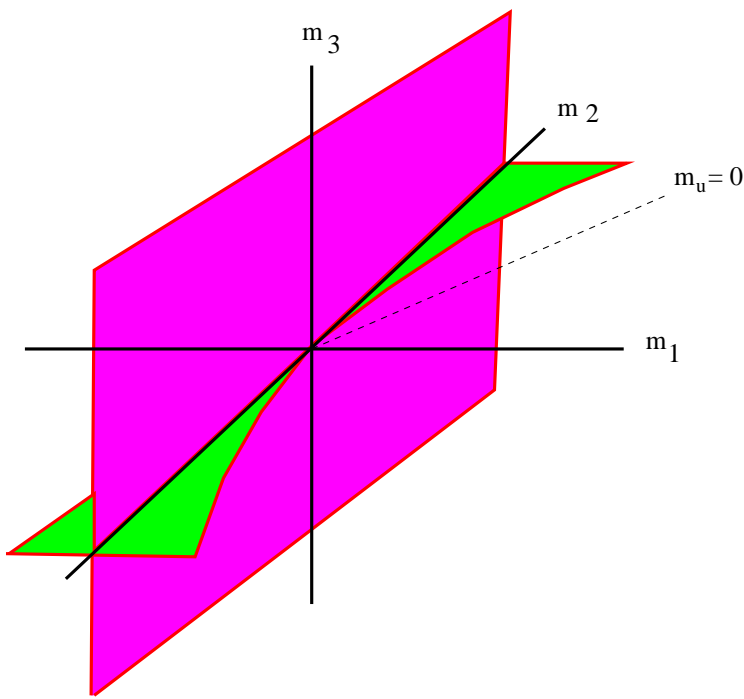}
\caption{The full phase diagram as a function of the three mass
  parameters.  
}
\label{fig6}
\end{centering}
\end{figure}

The final phase diagram as a function of the three mass parameters
appears in Fig.~\ref{fig6}.  There are two intersecting first order
surfaces, one at $(m_1=0$, $m_3\ne 0)$ and the second contained in the
region $(m_1<m_2$, $m_3=0)$.  The second surface ends along a critical
line.  In conventional language, these transitions all occur when the
strong CP angle takes the value $\pi$, but it is important to note
that there is a finite region with $\Theta=\pi$ without any phase
structure, {\it i.e.} when $m_2$ is only slightly larger than $m_1$.
Here the quark masses differ in sign, but one is much smaller than the
other in magnitude.

\vfill\eject

\section{Summary}

Non-perturbative effects can result in a mixing between the masses for
different quark species.  Because this effect is absent in
perturbation theory, it is inappropriate to match lattice and
perturbative calculations of quark masses, particularly when they are
non-degenerate.

The two flavor theory depends on three possible mass parameters.  One
of these is explicitly CP violating; its apparent absence is the
strong CP problem.  As these three parameters are varied from their
physical values, a rather rich phase diagram is encountered, displaying
both first and second order transitions.  In this diagram there is no
structure at $m_u=0$ when $m_d\ne 0$.  This is closely connected with
the result that $m_u=0$ is not an appropriate solution to the strong
CP problem.

\begin{thebibliography}{10}

\bibitem{Creutz:2009zq}
Michael Creutz.
\newblock {Anomalies and discrete chiral symmetries}.
\newblock {\em PoS}, QCD-TNT09:008, 2009.

\bibitem{Creutz:2010ts}
Michael Creutz.
\newblock {Quark mass dependence of two-flavor QCD}.
\newblock {\em Phys. Rev.}, D83:016005, 2011.

\bibitem{Creutz:2011hy}
Michael Creutz.
\newblock {Confinement, chiral symmetry, and the lattice}.
\newblock {\em Acta Physica Slovaca}, 61:1--127, 2011.

\bibitem{Georgi:1981be}
Howard Georgi and Ian~N. McArthur.
\newblock {Instantons and the $m_u$ Quark Mass}.
\newblock {\em unpublished (HUTP-81/A011)}, 1981.

\bibitem{Banks:1994yg}
Tom Banks, Yosef Nir, and Nathan Seiberg.
\newblock {Missing (up) mass, accidental anomalous symmetries, and the strong
  CP problem}.
\newblock {\em unpublished (hep-ph/9403203)}, 1994.

\bibitem{Creutz:2003cj}
Michael Creutz.
\newblock {CP symmetry and the strong interactions}.
\newblock {\em unpublished (hep-th/0303254)}, 2003.

\bibitem{Creutz:2003xc}
Michael Creutz.
\newblock {Ambiguities in the up quark mass}.
\newblock {\em Phys. Rev. Lett.}, 92:162003, 2004.

\bibitem{Bazavov:2009bb}
A.~Bazavov, D.~Toussaint, C.~Bernard, J.~Laiho, C.~DeTar, et~al.
\newblock {Nonperturbative QCD simulations with 2+1 flavors of improved
  staggered quarks}.
\newblock {\em Rev.Mod.Phys.}, 82:1349--1417, 2010.

\bibitem{Blum:2010ym}
T.~Blum, R.~Zhou, T.~Doi, M.~Hayakawa, T.~Izubuchi, et~al.
\newblock {Electromagnetic mass splittings of the low lying hadrons and quark
  masses from 2+1 flavor lattice QCD+QED}.
\newblock {\em Phys.Rev.}, D82:094508, 2010.

\bibitem{Durr:2010vn}
S.~Durr, Z.~Fodor, C.~Hoelbling, S.D. Katz, S.~Krieg, et~al.
\newblock {Lattice QCD at the physical point: light quark masses}.
\newblock {\em Phys.Lett.}, B701:265--268, 2011.

\bibitem{Frezzotti:2003ni}
R.~Frezzotti and G.C. Rossi.
\newblock {Chirally improving Wilson fermions. 1. O(a) improvement}.
\newblock {\em JHEP}, 0408:007, 2004.

\bibitem{Munster:2004wt}
Gernot Munster, Christian Schmidt, and Enno~E. Scholz.
\newblock {Chiral perturbation theory for twisted mass QCD}.
\newblock {\em Nucl.Phys.Proc.Suppl.}, 140:320--322, 2005.

\bibitem{Adler:1969gk}
Stephen~L. Adler.
\newblock {Axial vector vertex in spinor electrodynamics}.
\newblock {\em Phys.Rev.}, 177:2426--2438, 1969.

\bibitem{Adler:1969er}
Stephen~L. Adler and William~A. Bardeen.
\newblock {Absence of higher order corrections in the anomalous axial vector
  divergence equation}.
\newblock {\em Phys.Rev.}, 182:1517--1536, 1969.

\bibitem{Bell:1969ts}
J.S. Bell and R.~Jackiw.
\newblock {A PCAC puzzle: pi0 to gamma gamma in the sigma model}.
\newblock {\em Nuovo Cim.}, A60:47--61, 1969.

\bibitem{Jackiw:1999bc}
Roman Jackiw.
\newblock {What good are quantum field theory infinities?}
\newblock 1999.
\newblock hep-th/9911071.

\bibitem{Fujikawa:1979ay}
Kazuo Fujikawa.
\newblock {Path integral measure for gauge invariant fermion theories}.
\newblock {\em Phys.Rev.Lett.}, 42:1195, 1979.

\bibitem{Creutz:2010ec}
Michael Creutz.
\newblock {Anomalies, gauge field topology, and the lattice}.
\newblock {\em Annals Phys.}, 326:911--925, 2011.

\bibitem{Creutz:2007yg}
Michael Creutz.
\newblock {Chiral anomalies and rooted staggered fermions}.
\newblock {\em Phys. Lett.}, B649:230--234, 2007.

\end{thebibliography}

\end{document}